\documentclass{article}

\usepackage{arxiv}

\usepackage[utf8]{inputenc} 
\usepackage[T1]{fontenc}    
\usepackage{hyperref}       
\usepackage{url}            
\usepackage{booktabs}       
\usepackage{amsfonts}       
\usepackage{nicefrac}       
\usepackage{microtype}      
\usepackage{lipsum}		
\usepackage{graphicx}
\usepackage{url}

\title{Carbon Neutrality Approaches for IoT-Enabled Applications -- A Survey}

\author{ \href{https://orcid.org/0000-0002-2543-2710.}{\includegraphics[scale=0.06]{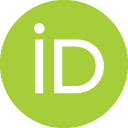}\hspace{1mm}Shajulin~Benedict}\thanks{(\url{www.sbenedictglobal.com}---\emph{© IEEE ICAISE 2023 Personal use of this material is permitted. Permission from IEEE must be obtained for all other uses, in any current or future media, including reprinting/republishing this material for advertising or promotional purposes, creating new collective works, for resale or redistribution to servers or lists, or reuse of any copyrighted component of this work in other works.}}) \\
	Faculty, Department of Computer Science\\
	Indian Institute of Information Technology Kottayam \\
	Valavoor P.O., Kottayam District, Kerala, India -- 686635. \\
	\texttt{shajulin@iiitkottayam.ac.in} \\
    \texttt{www.sbenedictglobal.com} 
}

\begin{document}
\maketitle

\begin{abstract}
	Unlike others, IoT-enabled technology has expanded its base in various sectors, including finance, healthcare, agriculture, energy, and so forth. Tens of thousands of applications and products have evolved in recent years, leading to a constant threat to global climate sustainability due to the underlying carbon emissions. Researchers have developed standalone methods/approaches that aimed to tackle carbon emission problems. However, an article that expresses a list of carbon-neutral solutions and their associated technological challenges is not available. This article explores the carbon emission problems of IoT-enabled applications; and, it categorizes the carbon neutrality methods. The exploratory analysis of different carbon-neutral approaches, presented in this article, can present novel ideas and answers to questions that hover over any carbon-conscious IoT-enabled application developers.    
\end{abstract}


\section{Introduction}
Applications surpassing from healthcare to financial touchless transactions utilize sensor-enabled devices and cloud services. The associated networked nodes have to be energy conscious so that carbon emissions are reduced -- one of the crucial social challenges. In fact, carbon emissions drastically impact the climate change \cite{carbon:2022:web} conditions of the globe. The emission constantly increased such that it reached over 50 billion tons of carbon emissions per year -- notably, the carbon emissions have reached 420 parts per million by 2022. 

Several regulatory bodies and tax levying mechanisms have been adopted in the recent past to neutralize such carbon emissions. For instance, the Paris agreement was introduced in 2015 to legally bind countries for neutralizing carbon emissions \cite{carbon1:2022:web}; similarly, a few observatory projects such as Atmospheric InfraRed Sounder (AIRS) from NASA\cite{carbon3:2022:web} and Orbiting Carbon Observatory from California Institute of Technology \cite{carbon4:2022:web} have been developed to measure carbon emissions and provide specific carbon neutral solutions. 

Although carbon emissions have become more prevalent in the energy, industrial, and transport sectors \cite{carbon5:2022:web}, the execution of IoT-enabled applications, especially compute-intensive data science algorithm-specific applications on cloud data centers or fog nodes have shown a substantial increase in energy consumption factors -- i.e., the introduction of cooling systems or allied methods have increased carbon emissions.  

IoT-enabled application developers or associated machine learning researchers are interested to neutralize carbon emissions. However, they encounter specific challenges as listed below: 
\begin{enumerate}
\item Non-Awareness -- researchers are not aware of reducing carbon emissions in their computing platforms. For instance, researchers would be interested in finding options to reach zero carbon emissions for their applications;
\item Policy Issues -- industries and their associated countries follow different policies to target carbon emissions. For instance, there are policies to promote carbon neutrality and zero carbon emissions. Such policies represent the scale of carbon emissions. Reducing carbon emissions at the industry-level or computing center level is one aspect of carbon emissions. However, availing the viewpoints of different governments to remove carbon-related gases at a higher scale is a non-converging policy-related issue. Additionally, researchers or practitioners are not aware of such policies or procedures; and,
\item Reduction Approaches -- providing carbon neutrality solutions or zero carbon enabling methods is at an infant stage, especially via. IoT-enabled applications. 
\end{enumerate}

\begin{figure*}[h]
\centering
\includegraphics[scale=0.4]{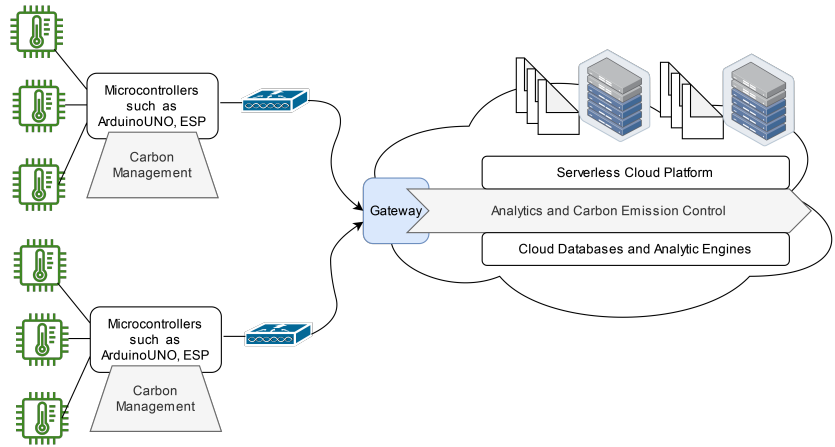}
\caption{Generic Architecture for IoT Cloud Applications -- Carbon Neutrality Perspective}
\label{fig:architecture}
\end{figure*}

This article pinpoints the research directions for IoT-enabled application developers or practitioners who prefer to incorporate carbon-neutral solutions. It delivers a taxonomy of mechanisms that apply to enabling carbon-neutral solutions in IoT cloud applications. In addition, the most predominant research directions that can be carried out by carbon-neutral conscious researchers in the near future are expressed in the article. 

The rest of the article is explained as follows: Section~\ref{sec:related} discusses the existing survey works relating to carbon emissions; Section~\ref{sec:taxonomy} highlights the taxonomy of carbon emission reduction techniques; Section~\ref{sec:analysis} compares various available research works and their pitfalls with suitable research directions; and, finally, Section~\ref{sec:conclusion} provides a few conclusions of this article. 

\section{Related Works} \label{sec:related}
Carbon neutrality has become a point of discussion in recent years owing to increased CO2/CO emissions across national borders. The carbon emissions are targeted in various sectors, including data centers and energy sectors. 

Efforts have been undertaken in the recent past to directly minimize carbon emissions by removing them from industrial wastes \cite{Arys:2017:Conf} or to indirectly minimize them by monitoring the emissions \cite{Peter:2022:Journal}. A few researchers studied the factors that lead to neutralizing carbon emissions. For instance, authors of \cite{Xiangshi:2021:Conf} studied the factors that influence carbon emissions in Pingdingsan city in China by developing a panel regression model; similarly, Maohui et al. \cite{Maohui:2022:Conf} studied the factors that influence carbon emissions focusing on Tianjin port city of China. 

Reducing carbon emissions targeting applications or associated computing nodes has been discussed by a few other researchers. For instance, modeling carbon emissions based on energy consumption of applications was discussed in \cite{Stefano:2022:Journal}; authors of \cite{Jesse:2022:Conf} developed a measurement framework to monitor the intensity of carbon emissions while executing machine learning models on cloud instances. Apart from controlling carbon emissions from applications, there have been efforts to protect carbon emissions from buildings or solid substances \cite{Xiou:2022:Conf} in a programmatic approach.

This article attempts to explore possible carbon emission reduction possibilities and neutralization factors while executing IoT-enabled applications on computing machines. 

NOTE: Read the final version for more details on related works. 

\section{Carbon Neutrality Approaches} \label{sec:taxonomy}
IoT-enabled applications have emerged in recent years in multitudes. Most of these applications involve learning processes from sensory software modules that provide text, video, or audio content. Processing these applications is not only compute-intensive but also network and memory intensive, leading to carbon emissions. This section explains a few procedures to target these issues after describing a generic IoT cloud architecture that attempts to neutralize carbon emissions. 

\subsection{Generic Architecture -- IoT Cloud Application}
IoT-enabled applications utilize sensors and actuators to capture measurable properties such as light, temperature, and so forth, and perform actions. All actions are combined with intelligence which involves machine learning algorithms such as Convolutional Neural Networks, Support Vector Machines, or Random Forests. 

\begin{figure}[h]
\centering
\includegraphics[scale=0.3]{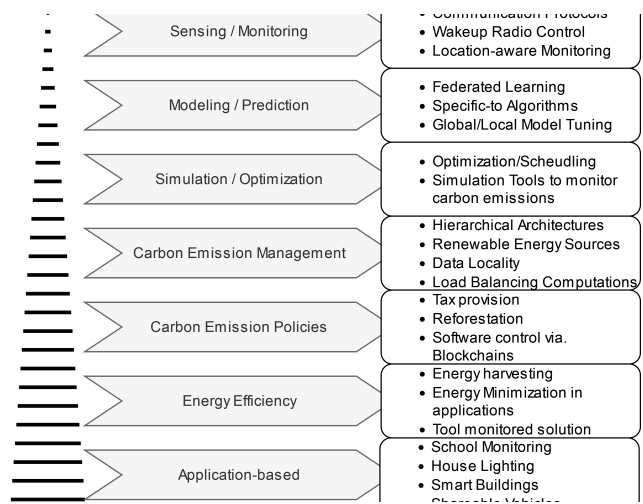}
\caption{Taxonomy of Research Works for Carbon Neutrality}
\label{fig:taxonomy}
\end{figure}

Articulating intelligence from sensed objects is often carried out using hierarchical layers of computing nodes as depicted in Figure~\ref{fig:architecture}. In recent years, to preserve privacy and to improve network latency, edge devices such as Jetson nano, Raspberry pi, and so forth, are utilized. Performing edge intelligence without fans contributes to minimizing carbon emissions rather than sharing the information over cloud/fog nodes. Also, powering off the computing devices and writing applications using serverless functions on serverless platforms increase energy efficiency. Obviously, edge-enabled hierarchical computing architectures are implemented in IoT-enabled applications. The carbon emissions on such IoT cloud architectures/frameworks could be neutralized or minimized using several techniques discussed in the following sections (see Figure~\ref{fig:taxonomy}). 

\subsection{Sensing and Monitoring}
One approach to handle carbon emissions is to sense carbon emissions using tiny to large-scale sensors \cite{Asthana:2018:Conf,Naser:2020:Journal}. This means that users could capture the air quality values of environments for preferring carbon-neutral locations. 

Based on the deadlines involved in sensing, the carbon emission sensing approach is classified into real-time and non-real-time sensing. For instance, in places where industries are located, it is a better choice to continuously monitor carbon emissions and to avoid any strange behaviours. On the contrary, locations, where the routine monitoring is required without deadlines, are done without any real-time considerations. 

In addition, while utilizing sensor-based applications, it is a better choice to enable carbon-efficient communication protocols such as Long Range Wide Area Network (LoRA) \cite{Hafiz:2021:Journal,Gilles:2021:Journal} or Message Queue Telemetry Transport (MQTT). On the contrary, if other communication protocols such as WiFi or 5G/6G networks are applied, the carbon emission of applications increases.  

\subsection{Modeling or Prediction}
Applying artificial intelligence (AI) in problem domains can reduce carbon footprints because such applications could be stopped from executions. In the past, a few researchers have developed mathematical models that revealed the emission of carbon footprints while implementing applications. For instance, authors of \cite{Kezban:2022:Conf, Sukhpal:2019:Journal} designed models that highlighted the carbon emissions from computing machines. In fact, these models reveal an indirect influence on the energy consumption of applications. 

In \cite{Kezban:2022:Conf}, authors developed a global model of carbon emissions using carbon detecting sensors and AI methods. 

\subsection{Simulation and Optimization}
Simulating the behavior of carbon emissions while executing IoT-enabled applications creates awareness among decision-makers or policy providers. In this lieu, a few simulators have been implemented. For instance, authors of \cite{Samreen:2015:Conf} developed a simulator to understand the level of carbon emissions. In real cases, the carbon emissions could be captured using satelite images or sensor networks. 

Scheduling jobs or optimizing them based on carbon emissions are specific approaches to neutralizing carbon emissions. Authors of \cite{Yucen:2017:Journal,Liang:2014:Journal} developed a Lyapunov algorithm to optimize jobs considering carbon emission; authors of \cite{Juan:2022:Journal} developed a genetic algorithm-based multi-objective algorithm to optimize resources; and, authors of \cite{Songling:2019:Journal} implemented ant colony optimization algorithm to identify suitable combinations/sequences to minimize carbon emissions. 

\subsection{Carbon Emission Management}
Apart from developing mathematical models, it is also a prevalent choice to predict carbon emissions of applications. Carbon emissions are managed in various means: 
\begin{enumerate}
\item Hierarchical Architectures -- As mentioned in the generic architecture section, emitting carbon dioxide can be minimized by involving hierarchical computing architectures. For IoT cloud applications, sensor nodes connect with varying levels of networked computing systems. These sensor nodes, especially when sensing video/images, transfer a huge volume of data to cloud services if there are no hierarchical computations. Also, while processing such applications on the respective compute nodes, they are not cooled if computed in edge nodes. 

\item Renewable Energy Sources -- Including more renewable energy sources while computing the intelligence of sensors will improve the neutrality of carbon emissions. Renewable energy sources are found abundant on our earth through sunlight, wind, water, waste, and so forth. However, traditionally, most of the computations are performed using fossil fuels such as coal, natural gas, or oil. 

\item Managing Data -- In computation, providing data immediately when the CPU requires them can minimize the underlying carbon emission of applications. In IoT-enabled applications, data are often available in geographically distributed cloud nodes. If an application is unable to locate the apt data or is not capable to offload data nearer to the compute nodes, they suffer from huge carbon emissions. In such cases, the memory/storage units have to be powered off. 

\item Load Balancing -- As similar to data localization, it is also crucial to minimize the computations of IoT-enabled applications in an efficient manner. For instance, while practicing federated learning tasks in a hierarchical architecture, it is an option to balance learning tasks among different computing nodes. In this way, some computations are handled by edge nodes, and a few other learning tasks are carried out by clouds. If a configuration of these learning algorithms is incorrect, it induces a large emission of carbon \cite{Xinchi:2022:Journal}. Obviously, carbon neutrality is possible through the development of a robust load balancer for IoT-enabled applications. 
\end{enumerate}

\subsection{Policies}
Fixing policies and practicing them can minimize carbon footprints. Many countries and industries have committed themselves to practice carbon neutrality by paying back compensations for any violations \cite{Jessie:2020:Journal}. It has also been accepted in many situations to collect taxes based on the utilization of carbon-related products that harm environments \cite{Huijuan:2017:Journal}. Also, ISO-controlled procedures have been implemented to promote carbon-efficient networked communications \cite{Julia:2021:Conf}. To neutralize emissions, another option is to plant trees. The number of plantations depends on the number of carbon emissions. To protect the enacted policies, it is better to include blockchains which can prevent the occurrence of data immutability issues, especially in corrupt organizations. 

\subsection{Energy Efficiency}
Reducing energy has a direct impact on carbon emissions. Hence, many researchers have attempted to improve the energy efficiency of applications, including IoT-enabled applications. In fact, several approaches have been practiced in the past for HPC and embedded applications. Notably, hardware-level monitoring using hardware counters and software-level energy monitoring of applications have been studied in \cite{Shajulin:2012:Journal}. Also, energy monitoring of edge devices with more emphasis on Expressif's ESP32 is studied in the past \cite{Oscar:2022:Conf}. 

Energy harvesting is a solution to utilize the available energy sources using sophisticated technologies. For instance, applying renewable energy sources from solar units for small IoT-enabled devices is an example of an energy harvesting mechanism. 

\subsection{Application-based}
At times, minimizing carbon emissions is practiced at the application-level. For instance, \textit{CO2 Meters} have been designed for monitoring carbon emissions in classrooms. The awareness of carbon emissions motivates people to lessen the emissions; Removing public lighting in streets tremendously can minimize the global carbon emissions of cities. An IoT-enabled solution that automatically controls the public lights based on renewable energy sources is one option to improve carbon emissions; An efficient building that considers natural lighting conditions can save carbon emissions at large scales; Shared mobility and shareable policies in the transportation sector can increase carbon neutrality features. The shareable vehicle can be supported by IoT-enabled solutions for increased cost efficiency and carbon efficiency. 

\section{Exploratory Analysis} \label{sec:analysis}
In this article, an exploratory analysis of various research works is carried out. To do so, research articles corresponding to the keyword ``carbon`` and ``IoT`` was searched using \texttt{scholar.google} and \texttt{Web-of-Science} databases. Among 560 downloaded articles, 230 research articles were filtered for further reading. Later, based on the scientometric analysis, a few most relevant articles that relate to IoT-enabled applications are picked up for explorations in two major directions: a) carbon management methods and b) carbon efficiency policies. In addition, in this section, a few research directions are highlighted for researchers/practitioners.

\begin{table}[]
\centering
\caption{Carbon Emission Management Methods in Literature}
\begin{tabular}{|l|l|l|l|l|l|}
\hline
  
\begin{tabular}[c]{@{}l@{}}Works\\ (Year)     \end{tabular} & Sector                                                    & Protocol                                                       & \begin{tabular}[c]{@{}l@{}}Learning/\\      Modeling\end{tabular} & \begin{tabular}[c]{@{}l@{}}Realtime/\\      Prediction\end{tabular} & \begin{tabular}[c]{@{}l@{}}Management/\\      Efficiency\end{tabular} \\ \hline
\cite{Hafiz:2021:Journal} (2021)                                                         & Energy                                                    & LoRA                                                        & Statistics                                                        & -                                                                   & \begin{tabular}[c]{@{}l@{}}Renewable\\      Energy\end{tabular}       \\ \hline
\cite{Gilles:2021:Journal} (2021)                                                         & \begin{tabular}[c]{@{}l@{}}Smart\\      City\end{tabular} & \begin{tabular}[c]{@{}l@{}}LoRA\\      NB-IoT\end{tabular} & -                                                                 & -                                                                   & \begin{tabular}[c]{@{}l@{}}Energy   \\      Harvesting\end{tabular}   \\ \hline
\cite{Badreddine:2019:Conf}     (2019)                                                     & \begin{tabular}[c]{@{}l@{}}Smart\\      Home\end{tabular} & MQTT                                                           & -                                                                 & Realtime                                                            & Hierarchical                                                          \\ \hline
\cite{Peter:2022:Journal}  (2022)                                                        & \begin{tabular}[c]{@{}l@{}}Power\\      Grid\end{tabular} & *                                                              & Deep RL                                                           & Prediction                                                          & AWS Cloud                                                             \\ \hline
\cite{Fan:2019:Conf}  (2019)                                                         & \begin{tabular}[c]{@{}l@{}}Smart\\      City\end{tabular} & Wifi                                                           & -                                                                 & -                                                                   & Cloud                                                                 \\ \hline
\cite{Xinchi:2022:Journal}   (2022)                                                        & Multi                                                       & Wifi                                                           & \begin{tabular}[c]{@{}l@{}}Federated\\      Learning\end{tabular} & \begin{tabular}[c]{@{}l@{}}Monitoring/\\      Realtime\end{tabular} & \begin{tabular}[c]{@{}l@{}}Data   \\      Partitioning\end{tabular}   \\ \hline
\cite{Stefano:2022:Journal}  (2022)                                                     & IIoT                                                      & Cellular                                                       & \begin{tabular}[c]{@{}l@{}}Federated\\      Learning\end{tabular} & \begin{tabular}[c]{@{}l@{}}Monitoring/\\      Analysis\end{tabular} & \begin{tabular}[c]{@{}l@{}}Load\\      Balancing\end{tabular}         \\ \hline
\cite{Jingpeng:2019:Journal} (2019)                                                         & \begin{tabular}[c]{@{}l@{}}Power\\      Grid\end{tabular} & -                                                              & \begin{tabular}[c]{@{}l@{}}Neural\\      Network\end{tabular}     & \begin{tabular}[c]{@{}l@{}}Realtime\\      Simulation\end{tabular}  & \begin{tabular}[c]{@{}l@{}}Energy   \\      Optimization\end{tabular} \\ \hline
\cite{Cui:2022:Conf}       (2022)                                                    & \begin{tabular}[c]{@{}l@{}}Smart\\      City\end{tabular} & -                                                              & \begin{tabular}[c]{@{}l@{}}Fast\\      DL\end{tabular}     & \begin{tabular}[c]{@{}l@{}}Realtime\\      Prediction\end{tabular}  & \begin{tabular}[c]{@{}l@{}}Energy   \\      Harvesting\end{tabular} \\ \hline
\cite{Hasan:2022:Conf}    (2022)                                                       & \begin{tabular}[c]{@{}l@{}}Smart\\      Home\end{tabular} & LoRA                                                              & \begin{tabular}[c]{@{}l@{}} - \\      \end{tabular}     & \begin{tabular}[c]{@{}l@{}}Realtime\\      \end{tabular}  & \begin{tabular}[c]{@{}l@{}}Monitoring   \\      Awareeness\end{tabular} \\ \hline

\end{tabular} \label{tab:emission}
\end{table}

\subsection{Carbon Emission Management Methods}
Carbon emission is managed in multiple approaches as discussed in Section~\ref{sec:taxonomy}. The implementation of novel hierarchical architectures that include renewable energy sources has been considered as the most practiced approach to enable carbon neutrality. Apart from this approach, load balancing and offloading approaches have been carried out in a few research works. 

Table~\ref{tab:emission} highlights a list of research works that have been carried out in the past to minimize carbon emissions with more emphasis on IoT-enabled applications. It could be observed that carbon emissions have been challenged in major sectors such as energy sector, smart city sector, home automation sector, industrial sector, and transportation sector. 

A few carbon emission management techniques such as partitioning data, balancing load across multiple compute machines, applying hierarchical computing approaches that involve edge-to-cloud computations, including energy harvesting techniques, and so forth, have been studied in these research works. For instance, \cite{Jingpeng:2019:Journal} applied the fast deep learning (DL) method and energy harvesting techniques. A few solutions have been designed for real-time implementations using microcontrollers. 

IoT-enabled applications, in general, utilize different communication protocols such as Bluetooth4.0, WiFi, LoRaWAN, Zigbee, and so forth. In the articles of consideration, the majority of the works attempted to minimize the energy consumption of applications using LoRA to indirectly minimize carbon emissions. For instance, works that focused on renewable energy sources or battery-operated sources utilized LoRA-based lightweight communication protocols \cite{Hafiz:2021:Journal,Gilles:2021:Journal,Hasan:2022:Conf}.  On the contrary, applications that focused on reducing carbon emissions based on learning approaches involved heavy computations or hierarchical architectures. In doing so, processing distributed learning algorithms or approaches such as federated learning mechanisms is efficiently handled considering the carbon emission patterns.  

\subsection{Carbon Neutrality Policies}
Enforcing carbon neutral policies across nations has become an important portfolio in recent years. Novel policies have been implemented and practiced to counter the emissions. Some policies have also been documented using sophisticated technologies. Table~\ref{tab:policies} throws some light on different carbon emission policies. Although some policies are much essential for controlling emissions, they have to be monitored in real-time to showcase the impacts to stakeholders, including the controlling authorities of nations. Taxing for the amount of carbon emissions is one unique approach that has been suggested in some countries. However, there are limited suggestions to tie up the findings with nation-level budget allocations. Also, committing to plant new trees based on the level of carbon emissions from applications is one aspect of enabling carbon neutrality.

\begin{table*}[]
\centering
\caption{Carbon Efficiency Policies and Allied Research Works}
\begin{tabular}{|l|l|l|l|l|l|l|}
\hline
  
\begin{tabular}[c]{@{}l@{}}Research\\      Work\end{tabular} & \begin{tabular}[c]{@{}l@{}}ISO-\\      Controlled\end{tabular} & Blockchain & \begin{tabular}[c]{@{}l@{}}Taxing\\      Capping\end{tabular} & Payback & Reforestation & \begin{tabular}[c]{@{}l@{}}Direct/\\      Indirect\end{tabular} \\ \hline
\cite{Michael:2020:Journal} (2020)                                                         & -                                                              & x          & -      & -       & -             & Concept                                                         \\ \hline
\cite{Huijuan:2017:Journal} (2017)                                                          & -                                                              & -          & x      & -       & -             & Direct                                                          \\ \hline
\cite{Robert:2021:Journal} (2021)                                                        & -                                                              & x          & -      & -       & -             & Indirect                                                        \\ \hline
\cite{Jessie:2020:Journal}  (2020)                                                      & -                                                              & -          & -      & x       & -             & Concept                                                         \\ \hline
\cite{Julia:2021:Conf}   (2021)                                                   & x                                                              & -          & -      & -       & -             & Direct                                                          \\ \hline
\cite{Amelia:2021:Conf}   (2021)                                                      & -                                                              & -          & -      & -       & x             & Concept                                                         \\ \hline
\cite{Sanabel:2022:Conf}   (2022)                                                       & -                                                              & -          & x      & -       & -             & Concept                                                         \\ \hline

\end{tabular} \label{tab:policies}
\end{table*}

\subsection{Research Directions}
Based on a detailed research analysis of available works, a few research directions are suggested for future researchers and practitioners who undertake their research in carbon-neutral IoT solutions. In this article, the research directions are focused on three major subtopics: 

\subsubsection{IoT-enabled Applications}
A few applications could be designed using sensors such that the architecture is powered using solar energy or other renewable energy sources. Several energy reduction mechanisms can be incorporated into sensor-based applications. Obviously, these mechanisms will have a direct impact on carbon neutrality. Especially, when IoT-based applications run for long durations, serverless-based cloud execution solutions may be incorporated. In the transportation sector, sensors may be powered to motivate car owners/drivers to switch to utilizing manual bikes. For instance, rewarding mechanisms may be implemented after suggesting a few emotionally-driven solutions. Additionally, lighting in homes could be controlled with power-saving LED lights that are attached to embedded edge devices.  

\subsubsection{Machine to Machine Support}
In the future, machines could decide on automatically protecting data and enabling actuators when involved with technologies such as blockchains. For instance, IoT-enabled solutions could be developed to automatically set budgets and thresholds by incorporating machine-to-machine communications. Notably, home heating and home cooling systems could be machine-controlled while setting up the electricity budgets that indirectly impact carbon emissions. 

\subsubsection{Assistive National Policies}
To control carbon emissions at a large scale, it is important to involve national-level policies. Usually, promoting newer national policies expects confidence from the majority of public members. This process could be assisted using social IoT techniques that involve sensing data from social media sites, including Facebook, Twitter, or Linkedin. 

\section{Conclusion} \label{sec:conclusion}
Neutralizing carbon emissions needs to incorporate modern techniques such as IoT-enabled monitoring approaches or energy harvesting methods. Application developers are not aware of available methods and carbon neutrality policies that could be adopted in their application domains. This article illustrated the possible carbon neutrality options either by sensing, modeling, simulating, or managing them. In addition, the most commonly applied policies to counteract carbon emissions have been discussed by comparing existing research works. The findings will be useful for researchers who undertake research directions based on IoT-enabled approaches for addressing carbon emissions. 



\end{document}